\documentclass[12pt,preprint]{aastex}

\newcommand{\drv}[2]{\frac{\partial #1}{\partial #2} }

\shorttitle{BXS survey - I}
\shortauthors{Ajello et al.}

\begin{document}

\title{BAT X-ray Survey - I: Methodology and X-ray Identification.}

\author{M. Ajello\altaffilmark{1}, J. Greiner\altaffilmark{1},
G. Kanbach\altaffilmark{1}, A. Rau\altaffilmark{2}, A. W. Strong\altaffilmark{1}, and J. A. Kennea\altaffilmark{3}}
\affil{$^1$ Max-Planck Institut f\"ur Extraterrestrische Physik, Postfach 1312, 
85741, Garching, Germany}
\affil{$^2$ Caltech Optical Observatories, MS 105-24, California, Institute of 
Technology, Pasadena, CA 91125, USA}

\affil{$^3$ Pennsylvania State University, 525 Davey Laboratory, UniversityPark, PA 16802, USA }

\email{majello@mpe.mpg.de}

\begin{abstract}
We applied the Maximum Likelihood  method, as an image 
reconstruction algorithm,
to the BAT  X-ray Survey (BXS). 
This method was specifically designed
to preserve the full statistical information in the data and to avoid 
mosaicking of many exposures with different pointing directions, thus
reducing systematic errors when co-adding images. 
We reconstructed, in the 14-170\,keV energy band,
the image of a  90x90 deg$^2$ 
sky region, centered on (RA,DEC)=105$^{\circ}$,-25$^{\circ}$,
 which BAT surveyed with an exposure time of
$\sim1$\,Ms (in Nov. 2005).
 The best sensitivity in our
image is $\sim0.85$\,mCrab or $2.0\times 10^{-11}$ erg cm$^{-2}$.
We detect 49 hard X-ray sources  above the 4.5\,$\sigma$ level; of these,
only 12 were previously known as hard X-ray sources ($>$15\,keV). 
Swift/XRT observations allowed us to firmly identify 
the counterparts for 15 objects, while 2 objects have Einstein IPC counterparts
\citep{harris90}; 
in addition to those,
we found a likely counterpart for 13 objects by correlating our sample with the
 ROSAT All-Sky Survey Bright Source Catalog \citep{voges99}. 7 objects
remain unidentified. 
Analysis of the noise properties of our image shows that $\sim75$\% of the
area is surveyed to a flux limit of
 $\sim$1\,mCrab.
This study shows that the coupling of the Maximum Likelihood method 
to the most sensitive, all-sky surveying, hard X-ray instrument, BAT, is able
to probe for the first time the hard X-ray sky to the mCrab flux  level.
The successful application of this method to BAT demonstrates that it could
also be applied with advantage to similar instruments like INTEGRAL-IBIS.
\end{abstract}

\keywords{galaxies: active -- surveys -- X-rays: binaries -- X-rays: galaxies}

\section{Introduction}
More than 40 years after its discovery, the nature of the Cosmic X-ray Background
(CXB) is still debated.
Population synthesis models,   based on unified AGN schemes, 
explain the CXB spectrum using a mixture of obscured and unobscured
AGN \citep[e.g.][]{comastri95,gilli01}.

According to these models, most AGN spectra are heavily absorbed, and
about 85\% of the radiation produced by super massive black hole
 accretion is obscured by dust and gas 
\citep{fabian99}.

Deep soft X-ray surveys (0.5--2.0\,keV) were able to resolve
the majority ($\approx 80$\%) of the 
CXB flux  into discrete sources \citep{hasinger98}.
However the resolved fraction decreases with energy, being $\sim$50-60\% 
in the 6--8\,keV band
\citep{giacconi02,rosati02} and even less above $>$8\,keV; the missing CXB component
has a spectral shape that is consistent with a population of yet undetected,
highly obscured AGN \citep[see][]{worsley05}.

It is important to realize that highly obscured objects are detectable 
in X-rays only above 10\,keV.
Moreover, most of the energy of the CXB
is emitted around 30\,keV \citep{marshall80} and the exact nature of the source 
population responsible for the background at these energies is unknown primarily
because of the low sensitivity of previous X-ray telescopes operating above 15\,keV.

All these reasons together motivate more sensitive observations of the hard X-ray sky.
\\\\
The Burst Alert Telescope  \citep[BAT;][]{barthelmy05}, on board the 
Swift mission \citep{gehrels04},  launched by NASA on 2004 November 20,
 represents
a major improvement in sensitivity for imaging of the hard X-ray sky.
BAT is a coded mask telescope with a wide field of view 
(FOV, $120^{\circ}  \times 90^{\circ}$ partially coded) aperture
sensitive in the 15--200\,keV domain.
BAT's  main purpose is to locate Gamma-Ray Bursts (GRBs).
While  chasing new GRBs, 
BAT surveys the hard X-ray sky  with an unprecedented sensitivity.
Thanks to its wide FOV  and its  pointing strategy, 
BAT monitors continuously up to 80\% of the sky every day.
Early results from the BAT survey \citep{markwardt05} show that BAT 
is already ten times more sensitive than the previous
hard X-ray all-sky survey performed by HEAO-1 \citep{levine84}.
\\\\
Coded mask telescopes are, until the advent of next generation hard X-ray 
focusing optics, among the most sensitive 
instruments able to image the sky in the hard X-ray domain.
Objects in the FOV cast part of the mask pattern onto the detector plane.
Since the sources' signal is coded by the mask onto the plane
this phase is also referred to as  coding phase. Thus,
a decoding procedure is required in order to reconstruct the original sky image.
A variety of methods can be used to reconstruct the sky image in the case 
of a coded mask
aperture \citep[see][for a general discussion on reconstruction
methods]{skinner87}.
 Among them, standard cross correlation of the shadowgram 
with a deconvolution array, the mask pattern, via FFT transforms, 
is the most often used.
Generally, sky images are obtained for each individual observation, 
where an observation is defined as a period during which the attitude is 
stable and constant. Subsequently, another procedure, 
such as resampling and reprojecting, is needed in order to assemble
the final all-sky image.
\\\\
Most of the extragalactic sources are very faint in the hard X-ray band. Thus 
their detection is challenging and requires sensitive techniques.
We here describe the application of an alternative 
method which was designed to improve the sensitivity avoiding
some of the disadvantages of the standard mask unfolding technique.
\\\\
This study has been performed in the framework of a campaign for optical spectroscopy
analysis of a sample of ``hard X-ray selected'' extragalactic  sources aimed
at identifying new Sy2 galaxies. This paper discusses the method 
used to reconstruct the survey image and presents the source catalog.
A second paper \citep{rau07}
describes in details the optical campaign and the source
identification process; the spectral analysis and the statistical
properties of the source sample are discussed in \cite{ajello07b}.
\\\\
The structure of the paper is as follows. In Section \ref{sec:ml} we present
details of the Maximum Likelihood method that was developed to analyze the BAT data.
In Section \ref{sec:analysis}, we describe the analysis steps performed and we present
and discuss the results of our image reconstruction algorithm. 
The last section summarizes the results. 
%%%%%%%%%%%%%%%%%%%%%%%%%%%%%%%%%%%%%%%%%%%%%%%%%%%%%%%%%%%%%%%%%%%%%%%%%%%%%%%%%
% Method Section

\section{Spatial Model Fitting} \label{sec:ml}

We apply ``spatial model fitting'', as described in  \cite{strong05}, 
to directly reconstruct the survey image from the raw detector data.
``Spatial model fitting'' means that a 
number of sky distributions, whose linear combination constitutes the model, 
are forward-folded through
the full instrumental response in order to generate a model  shadowgram.
The model shadowgram, which is a linear combination of all model components,
is then fitted in the full data space in order to get 
the most probable sky distribution.
The actual search for  unknown sources is then realized by moving a source probe 
   in a grid over the sky.
It is worth noting that no other steps, as image mosaicking, are required at
the end of this process. This method was successfully applied
to different kinds of experiments 
\citep[i.e. COMPTEL and INTEGRAL-SPI;][]{diehl95,strong05}. Its
development was driven by the capabilities 
of reducing systematic errors and noise e.g. the noise related with individual
short images and the systematics when co-adding noisy images 
in the mosaicking procedure.
This leads to an improvement in sensitivity over other methods, in particular 
reducing systematic errors from background variations and resampling. The full
information in the data is preserved and correctly treated in a statistical sense.
\\\\
The likelihood is the probability of the observed BAT data given the  model.
For our case it is defined as the product of the probability for each detector of each 
observation:
\begin{equation}
L = \prod_{ijk}\  p_{ijk}.
\end{equation}

where 
\begin{equation}
p_{ijk} = \frac{\theta_{ijk}^{n_{ijk}} e^{-\theta_{ijk} }}{n_{ijk}!}.
\end{equation}

is the Poisson probability of observing $n_{ijk}$ counts in pixel $ij$,
during the $k$-th observation,
when the number of counts predicted by the model is $\theta_{ijk}$.

The model is a linear combination of components; in the simplest case of
1 non variable source   and 1 background component for each observation, we get:
\begin{equation}\label{eq:model}
\theta_{ijk}= c_0\times(A\otimes S^{\alpha_0, \delta_0})_{ijk}+c_k \times B_{ijk}
\end{equation}
where $(A\otimes S^{\alpha_0, \delta_0})$ is the convolution of the detector
response ($A$) and a source of unit flux ($S$) 
at the  sky position $\alpha_0,\delta_0$
and thus $(A\otimes S^{\alpha_0, \delta_0})_{ijk}$ yields
the prediction of counts from a unit flux source
at the sky position $\alpha_0,\delta_0$ in detector $ij$, during observation $k$;
$B_{ijk}$ is the background prediction for pixel $ij$ in observation $k$ and 
$c_0$ and $c_k$'s are the parameters we want to estimate.\\

For the analysis described in this paper, the background model comprises, 
for each observation,  an empirical model (i.e. a 2 dimensional quadratic
function similar to the one used by the tool {\it batclean} 
as described in section \ref{sec:preproc}) and  the model shadowgrams for 
all bright sources (see also section \ref{sec:preproc}). 
The actual fit to the background is performed only
once; 
during the source search only the normalization of the background
in each pointing is allowed to vary.  Since sources detected at this
stage are faint, the background normalizations are expected
to vary by very small quantities. Indeed, 
we verified that such variations were less than 10$^{-3}$ with respect to
the background parameters determined before the source search.

In the future, our method will allow to test  more complex and 
physical background models (e.g. diffuse emissions).
%
% PARAMETER ESTIMATES
%
\subsection{Parameter model estimation}
The parameter values  are found by  maximizing the likelihood function, or, which is the 
equivalent, maximizing its logarithm:
\begin{equation}
\drv{\ln L}{\Lambda_i}= \drv{}{\Lambda_i}  \left[\sum_{ijk} n_{ijk} \ln(\theta_{ijk}) - \sum_{ijk} \theta_{ijk} \right].
\end{equation}

where $\Lambda$ is the vector of the parameters.
This translates into the following  set of equations:
\begin{equation}
\drv{lnL}{c_0} = \sum_{ijk} (A \otimes S)_{ijk} \left( \frac{n_{ijk}}{\theta_{ijk}} -1\right),
\end{equation}
\begin{equation}
\drv{\ln L}{c_k} =  \sum_{ijk}  B_{ijk} \left(    \frac{n_{ijk}}{\theta_{ijk}} -1       \right),\ \forall k 
\end{equation}
which allows to estimate all parameters simultaneously.\\
This set of equations can be solved only numerically and we use
a modified Newton algorithm in order to find the solution.
\subsection{Source significance}
In the case of a single source component, 
the source significance can be estimated using
the likelihood-ratio test. For this application, the null hypothesis is that no point
source exists at the position under consideration and the background model can explain
all the data. The alternative hypothesis is the converse.
Two maximizations have to be done in order to calculate the likelihood $L_0$ of the
background (null hypothesis) and the likelihood of both source and background for
the alternative hypothesis $L_1$.
The test statistic:
\begin{equation}
T_{\rm s} \equiv -2( \ln L_0 - \ln L_1)
\end{equation}

is expected, from  Wilks's theorem \citep{wilks38}, to be asymptotically 
distributed as $\chi^2_{n}$  in the null hypothesis, where $n$ is the
additional number of free parameters that are optimized for the alternative 
hypothesis. Since in our case the source intensity is the only additional free
parameter, the test statistics is expected to follow the $\chi^2_{1}$ 
distribution.
Thus, the significance of a detection can be addressed as:
\begin{equation}
S= \int^{\infty}_{T_s} \frac{1}{2} \chi^2_1(x) dx ,
\end{equation}
which, after changing variables, become:
\begin{equation}\label{gauss}
S= \int^{\infty}_{\sqrt{T_s}} \frac {e^{-y^2/2} }     {(2\pi)^{1/2}} dy .
\end{equation}
Equation \ref{gauss} is exactly the integral of the standard normal distribution
from $T_s^{1/2}$ to $\infty$ and so the significance of the detection is:
\begin{equation}
 T_s^{1/2}\sigma = \sqrt{ -2( \ln L_0-\ln L_1)}\sigma = n\sigma.
\end{equation}
Hence, by definition, the significance fluctuations must be distributed as a
normal Gaussian if everything is done correctly.
\subsection{Method implementation}
In case of large detector counts, the likelihood maximization is equivalent
to the $\chi^2$ minimization, 
 with the $\chi^2$ problem having the advantage 
that it can be solved faster analytically.
We have verified that  in the case
of large detector counts ($\geq$20) and large numbers of observations ($\geq$100)
the two solutions are very similar and from now on we use the $\chi^2$ solution.
\\\\
The algorithm used is a parallelized implementation  
of {\itshape spidiffit} \citep{strong05} used for INTEGRAL-SPI data analysis.
Parallelization was needed because of the size of the problem we are dealing with.
 The typical execution time needed to compute the analytical
$\chi^2$ solution scales with n$^2$ where n is the number of data points to fit
(i.e.: number of BAT detectors, 32768, multiplied by the number of observations).
A single minimization with 2600 observations takes nearly 90\,s; the total execution
time  to generate a map of 450$\times$450 pixels would be $\sim$200\,days.
This time has been reduced to $<$15 days using an average of 15 CPUs. We 
remark also that it is  the first time that such an approach is applied 
to a problem of this large size.
\\\\
As shown in equation \ref{eq:model}, the model is a 
linear combination of different components which can be specified at the input
of the program. Source and background components are in general treated in different
ways. Sources are assigned a single free parameter (their average intensity) 
while, as already discussed, the background components are allowed to vary
from pointing to pointing. However, in case of variable sources, the user
can specify that the source intensity is left as a free parameter in all
pointings (or in time-contiguous groups of them).
\\\\
We remark that for the analysis presented in the next sections, 
the program has been used
in its simplest configuration, with only one constant source and normalizations
of the per-pointing backgrounds allowed to vary.
However, after the source search had been performed and source candidates identified we have used the ability to  fit simultaneously all sources
(each of which was again assumed constant in time).
In fact, the simultaneous fit of all sources yields 
the best parameters (significances and fluxes)  
and allows us to discard spurious detections.
When the analysis is based on a large
number of observations correlation (``cross-talk'') between sources 
is negligible.

\subsection{Instrumental response}
As shown in equation \ref{eq:model}, the first part of the model represents the source
component (or components if more than one)  and this is  given in the most simple
form  by a 
point like source at position $\alpha_0,\delta_0$ in the sky, forward-folded
with the instrumental response.
We have used a large set of Crab observations ($>$ 1000)
to develope a parametrized diagonal full 
instrumental response which enables us to predict the expected counts 
(essentially the term $A\otimes S^{\alpha_0,\delta_0}$ of equation \ref{eq:model})
from a unit flux source as a function of energy and position in the 
field of view. 
The parametrized instrumental response was obtained in the following way
(standard BAT software is reported in brackets):
\begin{enumerate}
\item for each Crab observation a  model shadowgram for the source 
position is computed (tool {\it batmaskwtimg});
\item for each Crab observation the source model and the standard 
background components of {\it batclean} (see Section \ref{sec:preproc})
are fit to the data;
\item the normalizations of the source components (in the different observations)
are parametrized as a function of off-axis angle.
\end{enumerate}
The parametrized instrumental response is thus, 
for a given source position,
the multiplication of the model shadowgram described in (1) and of a coefficient
computed from the parametrization derived in (3). In this way the instrumental
response accounts for the off-axis\footnotemark{}
 variation of the detected source intensity
which the {\it batmaskwtimg} model does not take into account.
\footnotetext{The reader can find more details about the off-axis variation of 
the source signal and other effects in the   Appendix A.1 of \cite{ajello07b} }
The response, derived in this way, agrees with measured values 
to within 1 sigma anywhere in the FoV.

\begin{figure}[h!]
\begin{centering}
	\includegraphics[height=60mm,scale=1,width=70mm]{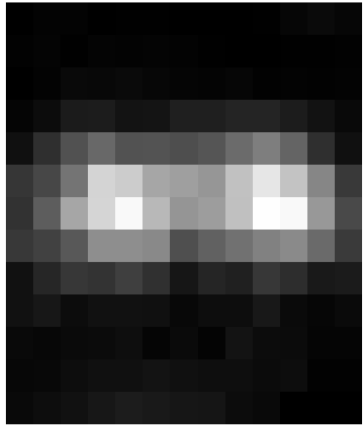} 
	\caption{Image of LMC X-1 (right) and PSR B0540-69.3 (left)
 clearly separated.  The pixel size is 5\arcmin\ and the map is about 
$1\times 1$ deg$^2$.}
	\label{fig:lmc}
\end{centering}
\end{figure}

To improve the speed of the code during source search
the full instrumental response was pre-computed
over a 6\arcmin\ pitch grid in  the whole BAT FOV.\\
\\\\
To conclude, in  Fig.~\ref{fig:lmc} we show 
the imaging reconstruction capabilities of our approach;
two closeby faint sources, LMC X-1 and PSR B0540-69.3, are clearly detected 
in the image obtained using $\sim$2600 observations. 
The good angular resolution of BAT is also preserved by
our imaging reconstruction algorithm, in fact the two sources are separated by just 
25\arcmin\ (for comparison the BAT Point Spread Function  is 22\arcmin).

%############################################################################
% ANALYSIS
\section{Analysis}\label{sec:analysis}

In this section we describe the application of the Maximum Likelihood method
to reconstruct the image of $\sim1/8$ of the sky using 8 months of BAT 
data.

% PREPROCESSING
\subsection{Data selection and screening}\label{sec:preproc}
We used 8\,months of data, from April 2005 (when BAT data became public) 
to November 2005. In order to secure optical follow-up with a dedicated 
observing campaign at La Silla, Chile in January 2006, we selected only 
observations
with angular separation less than 45\,degrees from the zenith 
(RA=105\,degrees, DEC=$-$25\,degrees).
The all-sky analysis, still within
the capabilities of modern super-computers, will be left to a future study.

Swift-BAT survey data are in the form of 80 channels detector plane 
histograms (DPH) with typical exposure time of  300\,s.\\
In order to have a suitable clean dataset as input of the imaging
reconstruction algorithm described in section \ref{sec:ml},
preprocessing  must be carried out on the raw survey data. 
This preprocessing phase accomplishes two different goals: 1) 
data quality is monitored along 
the processing and 2) the very bright sources detected during each single observations 
are localized and inserted in the background model
of the imaging reconstruction algorithm.
The latter procedure can be justified as follows.
The brightest sources (except the Crab Nebula)
are known to be highly variable. However, since they are detected in general
at high significance in a single observation, 
their intensities can be determined with good accuracy.
Thus, inserting bright sources in the background model, 
rather than treating them 
as several independent components in the source model, allows to handle
source variability in a natural way without increasing the size of the 
problem.

All the pre-processing was carried out using the  latest
available version of the Swift software contained in the 
HEASOFT 6.0.3. Below we report in brackets the name
of the standard BAT tools used during our pre-processing.
  
For each DPH, our pre-processing pipeline,  does the following operations:
\begin{enumerate}
\item{data are rebinned in energy channels according
to the gain-offset map generated on board ({\it baterebin})};
\item{the DPH is integrated along the energy axis,  between 14 and 170\,keV,
 and a detector plane image (DPI) is generated ({\it batbinevt})};
\item{a detector quality mask is created,
 where hot and cold pixels are masked out ({\it bathotpix})}. These pixels
are identified as the wings of the distribution of counts for a given 
observation; in general 2\% (and so roughly 1\% on each side)
of the distribution is excised;
\item{an empirical background model is fitted to the DPI
({\it batclean}\footnote{the empirical background model
 built-in in batclean fits (for a given energy range) a quadratic spatial
function plus a series of models which take care of detector edge effects for a 
total of 14 parameters.  The user is also free to include  sources or different  background models. The reader can find more details about the batclean background model
in the documentation included in the HEASOFT package (a copy of it is also 
available online at
 http://heasarc.nasa.gov/lheasoft/ftools/headas/batclean.html.)}});

\item{the DPI and the background model are input 
to a FFT deconvolution algorithm 
 which generates the sky image ({\it batfftimage})};
\item{source detection 
takes place on the sky image and a catalog of all sources detected
above S/N$>6\,\sigma$ is created ({\it batcelldetect}); }
\item{a model for each detected source 
is created and it is added to the background model of step 4. 
The source model is created using the measured source coordinates.
These coordinates were preferred to the catalog position because
of non-trivial systematic effects  which 
produced a shift in the measured source coordinates 
as a function of position in the FOV 
(see http://swift.gsfc.nasa.gov/docs/swift/analysis/bat\_digest.html for more
details) ; 
}
\item{steps 4 to 7 are repeated until no new sources are detected  in a 
single 300\,s observation.}
\end{enumerate}

In order to have the cleanest dataset possible we have applied cuts on the quality
of the data. During the steps above data are screened on the basis 
of the following conditions:
\begin{itemize}
\item{lock of the star tracker and pointing stability}
\item{spacecraft being outside of the South Atlantic Anomaly (SAA).
This information is reported in the housekeeping data and is referred
to a fiducial point inside the SAA.}
\item{BAT array rate $<$18000\,counts s$^{-1}$}
\item{exposure being larger than 200\,s}
\item{reduced $\chi^2$ of the background fit $<$1.5 }
\item{$>$ 9$\sigma$ detected sources must be within a distance of 0.1\,deg
 from a known source otherwise they are thought to be spurious or transient. 
The observation is flagged for a later analysis, but not inserted into the final dataset.}
\end{itemize}

In table~\ref{tab:data_frac}, we have listed the fraction of exposure
which is rejected if a single data quality cut is applied to the data
used for this analysis. For the current dataset, $\sim$ 34\% of the overall
exposure time was rejected because the data  did not meet one or more 
of the above mentioned criteria.

\begin{figure}[h!]
\begin{centering}
	\includegraphics[scale=1.0,trim= 0 0 0 20, clip=true]{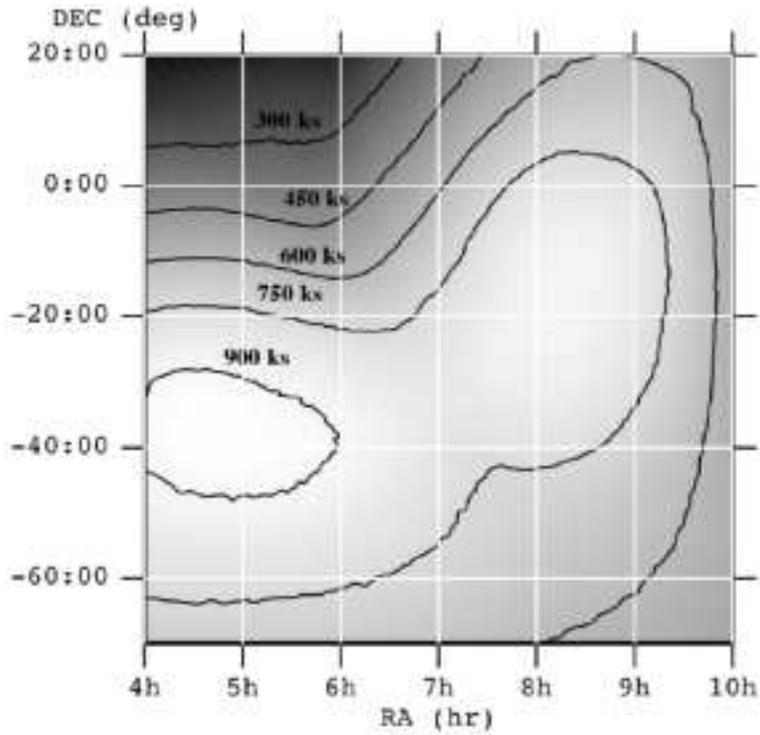} 
	\caption{Exposure map, corrected for telescope vignetting,
 of the survey field presented in text after data screening.
}
	\label{fig:exp}
\end{centering}
\end{figure}

After processing and screening the data according to these criteria, 
the final data set includes 2671 observations.
These observations are input to the imaging reconstruction algorithm 
described in section \ref{sec:ml}.  Fig.~\ref{fig:exp} shows
the total exposure map of all pointings.

\begin{figure}[h!]
\centering
\includegraphics[scale=0.7]{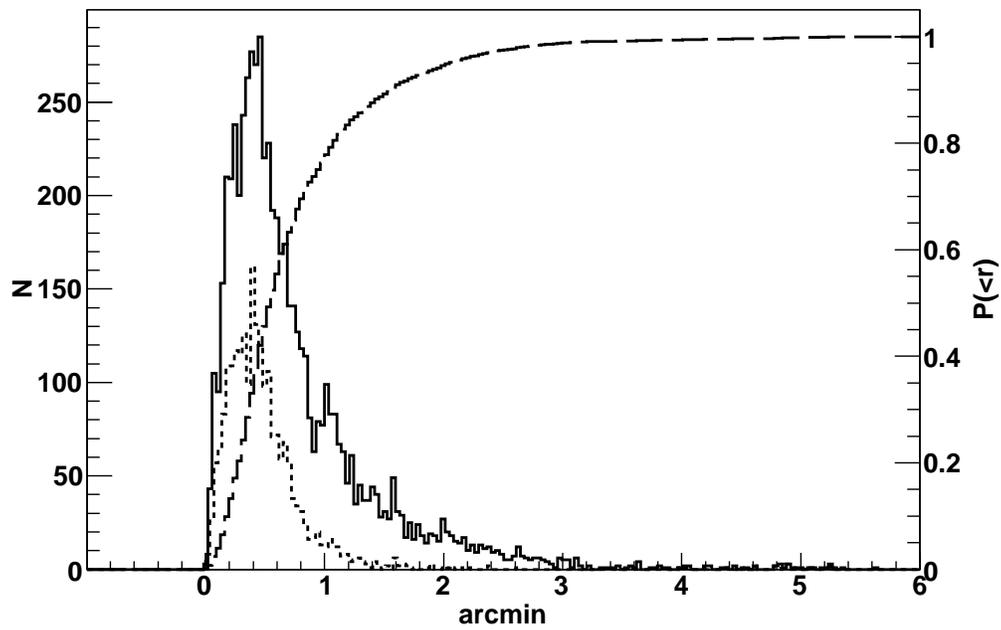} 
%\epsscale{0.7}
%\plotone{Offset_5min.eps}
	\caption{The solid line is the distribution of the offsets 
of sources detected in the individual observations (see  
Tab.~\ref{tab:prec_src}) from their catalog position. 
The inner histogram (dotted line) shows
the detections of the Crab  while the dashed line is the 
cumulative distribution of all detections.
 90\% and 95\% confidence limits are at  
radii of 1\hbox{$.\!\!^{\prime}$8} and 2\hbox{$.\!\!^{\prime}$2}  respectively.}
	\label{fig:off5min}
\end{figure}

\begin{figure}[h!]
\centering
\includegraphics[scale=0.7]{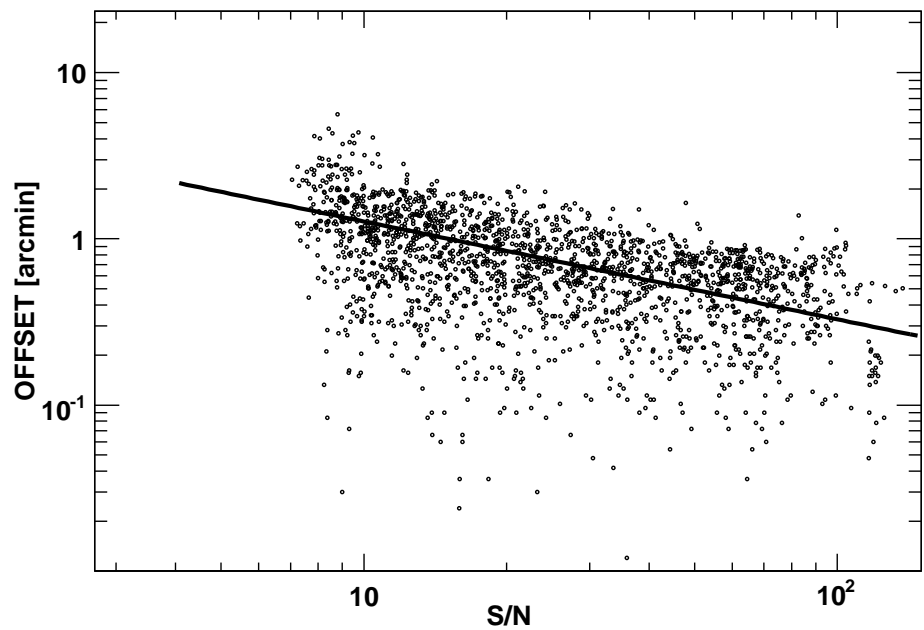} 
	\caption{Offset from catalog position 
of  sources detected in individual observations as
a function of S/N. The solid line is the function described in Eq.~\ref{eq:off}}
	\label{fig:off5min_sn}
\end{figure}

All sources detected 
during the preprocessing phase are listed in table~\ref{tab:prec_src} 
along with their identification, their maximum and total significance 
(computed as the sum of the squared of significances) 
from this per-pointing analysis, and the number of detections. 
The distribution of the offsets of sources in table~\ref{tab:prec_src} from their catalog 
counterpart is reported in Fig. \ref{fig:off5min}. The same graph shows the extremely
good location accuracy of BAT which locates 95\% of all  
sources, detected in single pointings, within 
2\hbox{$.\!\!^{\prime}$2} radius. \\
In order to understand the dependence of location accuracy
 on the source significance, 
we have analyzed  all per-pointing detections, see Fig.~\ref{fig:off5min_sn}, 
and determined that the 
%Crab Nebula detections and determined that the 
offset varies with significance accordingly to 
\begin{equation}
\textup{OFFSET}= 4.94(\pm 0.68 )\times (\textup{S/N})^{-0.59 (\pm0.05)}\ \ \ \ {\rm [arcmin]}
\label{eq:off}
\end{equation}

This analysis is based only on sources detected during 
individual  pointings.

%
%
%         IMAGE RECONSTRUCTION
%
%
\subsection{Imaging reconstruction}

\begin{figure}[h!]
\begin{centering}
\includegraphics[scale=1.3,trim=0 0 0 18,clip=true]{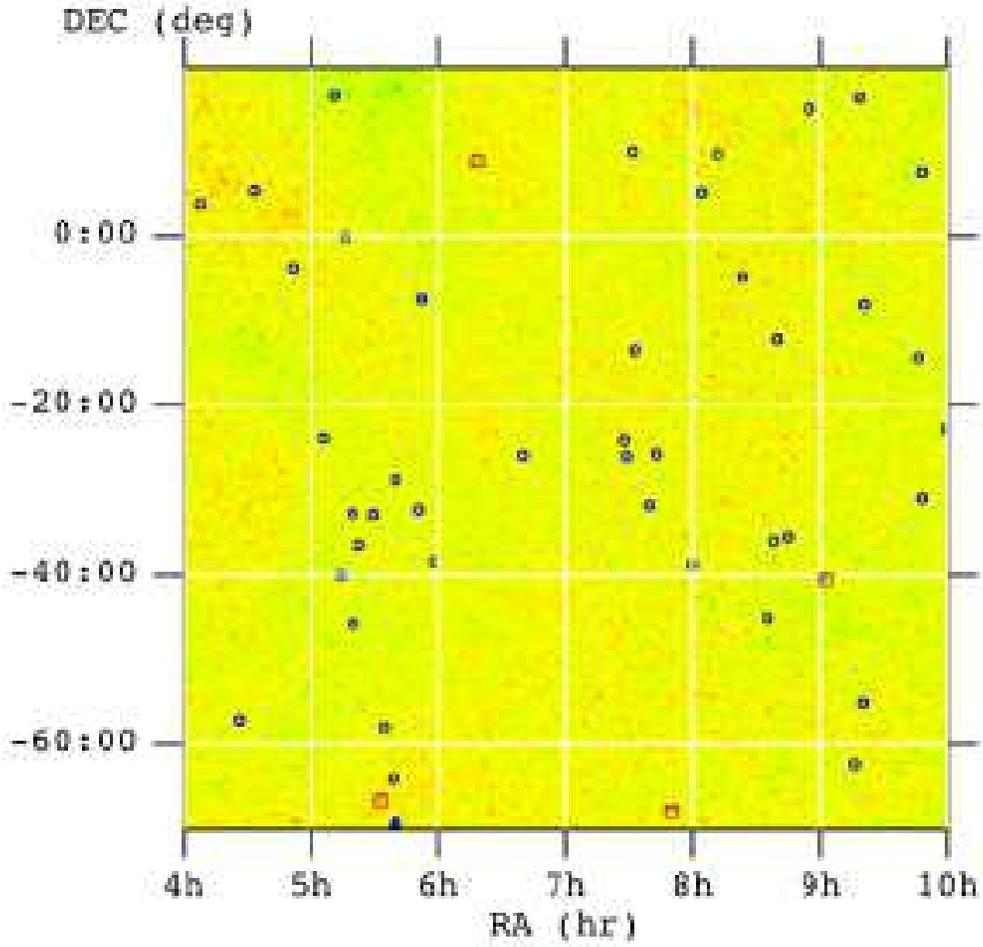}
\caption{Significance map of the 90x90\,deg$^2$ surveyed area. 
Blue circles mark all sources above 4.5\,$\sigma$ presented in 
Tab.~\ref{tab:src}, 
while red squares show the position of bright sources which were included
in the background model.}
\label{fig:map}
\end{centering}
\end{figure}

The 2671 DPIs along  with their background  models (created at step 4 in section 
\ref{sec:preproc}) are input of the imaging reconstruction algorithm. For this
analysis we have used 1 parameter for the source component; moreover 
we have allowed the
normalization of the background component to vary separately
in each pointing, leading to a total of 2672 parameters. 
The map is built in small segments of 5$\times$5 deg$^2$.
A pixel size of 12\arcmin\ was chosen as the best
compromise between computational time and the resampling factor of the PSF 
($\sim$ 2 in this case). The significance image is shown in Fig. \ref{fig:map}.\\

%
%%%%%%%%%%%%%%%%%%%%%%%%%%%%%%%%%%%%%%%%%%%%%%%%%%%%%%%%%%%%%%%%%%%%%%%%%%%%%%%%%%
%          Setting the significance threshold
%
\subsection{Setting the significance threshold}

There are several approaches in order to derive the best 
significance threshold.
The Maximum Likelihood method leads to perfectly symmetric Gaussian, normal,
noise in the pixels of the reconstructed image.
Thus,   the most straightforward approach is setting the threshold as the 
absolute value of the lowest negative fluctuation.
In this case, since the negative fluctuations are given by noise, one should expect
no false detection above this threshold.

\begin{figure}[h!]
\epsscale{0.7}
\plotone{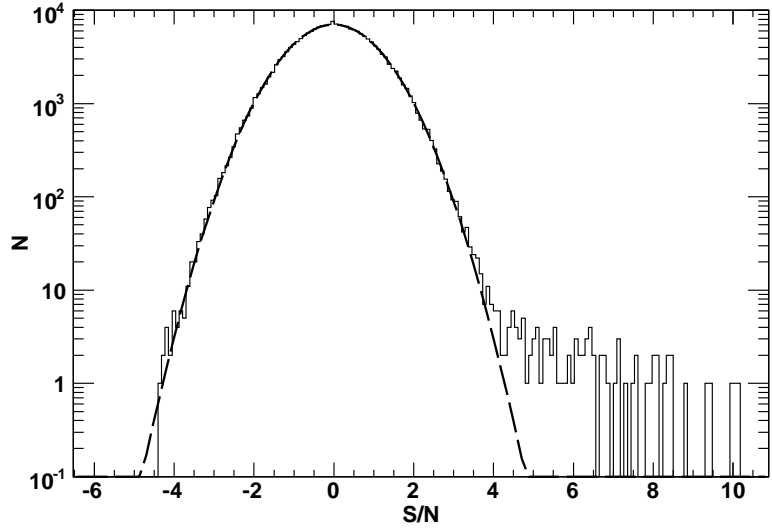}
\caption{S/N distribution.The dashed line is an overlaid
 Gaussian with $\sigma =1$.}
	\label{fig:snrdistr}
\end{figure}

As it can be seen in the  significance distribution reported in
Fig. \ref{fig:snrdistr}, no negative fluctuations larger than $-4.3$ are found.
If we take into consideration the number of trials and the  normal Gaussian 
distribution we get that  above the threshold S/N ratio of 4.5
we expect a number 
of false detections of 0.7. 
We also made a Monte-Carlo simulation generating a large number
($>$1000) of sky images with  Gaussian noise. We then counted
all the excesses above the 4.5\,$\sigma$ level and found out that
the number of expected false detection is 1.01, in agreement
with  the previous finding.  
A contamination of our sample of sources
by $\sim$1 spurious detection was judged to be a good compromise between
detection sensitivity and sample corruption (see Section \ref{sec:det}  for the 
chance connected to have a higher contamination).
Hence we decided to fix the  threshold to  4.5\,$\sigma$.

%
%%%%%%%%%%%%%%%%%%%%%%%%%%%%%%%%%%%%%%%%%%%%%%%%%%%%%%%%%%%%%%%%%%%%%%%%%%%%%%%%%%
%          Sky Coverage
%
\subsection{Noise properties and Sky coverage}

The sky coverage is, for a given survey, the distribution of the survey's area
as a function of limiting flux. The knowledge about the sky coverage is 
particularly important when computing the number-flux relation 
(also known as LogN-LogS distribution). We leave the derivation of the number-flux
relation to  a separate paper \citep{ajello07b}, but we are interested in deriving
the sky coverage here as it brings crucial information about the sensitivity
and noise properties of the survey.
\\\\
The sky coverage as a function of the minimum detectable flux $F_{min}$ is defined as
the sum of the area covered to fluxes $f_i < F_{min}$:
\begin{equation}
\Omega(<F_{min}) = \sum_i^N A_i \ \ \ \ f_i< F_{min}
\label{eq:skycov}
\end{equation}
where N is the number of image pixels and $A_i$ is the area associated with each
of them.

We have followed two procedures to compute the sky coverage of our survey area:
\begin{itemize}
\item the ML method produces a flux map and an error map as output of the 
fitting procedure.  In order to get the sky coverage we multiplied the error
map by the 4.5\,$\sigma$ threshold S/N ratio  and then counted the area as in Equation 
\ref{eq:skycov}; \\
\item  we computed the local (flux) image variance
using a sliding annular region whose internal and external radii
were 5 and 30 pixels respectively. The noise of a given pixel is thus
computed as the variance of the pixels contained in the annulus centered on it.
The central pixels are excised so that the background does not include
contamination from the source region. This map is a true 
representation of the noise in our image. Again, we multiplied this noise map
by our detection threshold of  4.5\,$\sigma$ and then counted the area 
as in Equation  \ref{eq:skycov}. 
\end{itemize}

The sky coverage computed in both ways does not present any significant differences
testifying that the error computed by the ML method is very close to
(if not the same as) the real noise term of the sky image.
In the left panel of Fig. \ref{fig:noise} we report the sky coverage of the entire area
and for the extragalactic portion of the sky (selected imposing 
$\mid b\mid>15^{\circ}$). As it can be seen from the sky coverage, $>$75\% of the
surveyed area is sensitive to fluxes $\sim$1\,mCrab and all of it to fluxes
$>$2.0\,mCrab. The limiting sensitivity in our image is a bit less than 
0.9\,mCrab  (or 2.05$\times 10^{-11}$erg cm$^{-2}$s$^{-1}$). 
\\\\
The analysis of the pixel noise as a function of exposure time
(reported in the right panel of Fig.~\ref{fig:noise})
shows that the survey sensitivity scales $\propto T^{-0.5}$ denoting that
systematic errors do not dominate over statistical ones. 
We then compared our survey sensitivity to
recent results from the BAT and INTEGRAL-ISGRI hard X-ray surveys 
\citep{markwardt05,
bassani06}. In order to perform the comparison, we transformed the sensitivities
provided by the authors in different bands to sensitivities in a common band 
(20-100\,keV); the comparison, which is shown in table \ref{tab:sens}, is done
in two ways: once taking into account the threshold S/N used by the authors 
in their work, and then also based on a common 5\,$\sigma$-equivalent sensitivity.
The main
result is that for 1\,Ms of  exposure, our survey is one of the most sensitive.

%
%%%%%%%%%%%%%%%%%%%%%%%%%%%%%%%%%%%%%%%%%%%%%%%%%%%%%%%%%%%%%%%%%%%%%%%%%%%%%%%%%%
%          Source detection and fluxes
%
\subsection{Source detections and fluxes}\label{sec:det}
Source detection on the reconstructed image is a straightforward process 
since  
significance and flux maps  are direct results of our reconstruction algorithm.
All not-neighbouring pixels which meet the criterion   
$S/N > S/N_{threshold}$ are identified. However at this stage we have
lowered our detection threshold to (an optimally chosen) 3.5\,$\sigma$.
Indeed, our procedure of using a pre-computed response over a 6\arcmin\ pitch
grid on a  12\arcmin\ pixel-size map might produce a small loss
in the reconstructed sources' fluxes and significances. 
In order to overcome this problem,
we have generated, for all the above candidates,
a 5\arcmin\ pitch grid map using the correct instrumental response 
(i.e. not pre-computed on a 6\arcmin\ pitch grid).
One such map has already been shown in Fig.~\ref{fig:lmc}. Only those
candidates
whose significance, as derived from the oversampled small maps, exceeds 
the 4.5\,$\sigma$ threshold  are kept in the sample and  fit with  
the instrumental point spread function  
(the {\it batcelldetect} is used here) in order to determine the most
accurate source parameters. 
This procedure allows us to recover the correct 
source significance and flux at the cost of a slightly larger number of false
detections. Indeed, due to the increased number of trials 
the expected number of false detection is now 1.5.
We remark that our map is one realization over many; 
thus there exists a not-zero probability that the number of false
detection exceeds the (averaged) value estimated here. Our Monte Carlo
simulation shows that the probability of getting
a number of false detections of 2, 3, and 4 is respectively
0.21, 0.09, and 0.02. 

% --------------------------------------
% these are the prob ==n
% Prob of =1 spurious source is 0.363000
% Prob of =2 spurious source is 0.210000
% Prob of =3 spurious source is 0.093000
% Prob of =4 spurious source is 0.033000
% Prob of =5 spurious source is 0.007000
% Prob of =6 spurious source is 0.001000
% Prob of =7 spurious source is 0.000000

% --------------------------------------
% these are the prob >= n
% Prob of >1 spurious source is 0.707000
% Prob of >2 spurious source is 0.344000
% Prob of >3 spurious source is 0.134000
% Prob of >4 spurious source is 0.041000
% Prob of >5 spurious source is 0.008000
% Prob of >6 spurious source is 0.001000
% Prob of >7 spurious source is 0.000000
% --------------------------------------

%%%%%%%%%%%%%%%%%%%%%%%%%%%%%%%%%%%%%%%%%%%%%%%%%%%%%%%%%%%%%%%%%%%%%%%%%%
% Detected Source
%
%
\subsection{Detected sources}
\begin{figure}[h!]
\begin{centering}
\includegraphics[height=90mm,scale=1,width=80mm,angle=270]{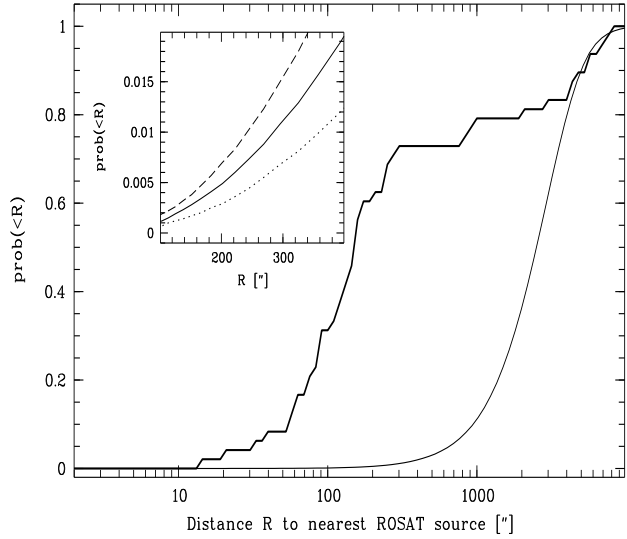}
\caption{Probability of finding at least one ROSAT source within a given radius
for our source sample of table \ref{tab:src}. The thick solid  line is the probability
for the real sample while the thin line is the expected contribution of chance
coincidences. The inset shows a closeup view of the chance coincidence curve
for the region of interest (200\arcsec\ $<$ R $<$300\arcsec); the dashed line 
is the chance coincidence distribution for latitudes
 $-40^{\circ}<$b$<-20^{\circ}$, while the dotted line
is for $0^{\circ}<$b$<20^{\circ}$. }
	\label{fig:ros_corr}
\end{centering}
\end{figure}

\begin{figure}[h!]
\begin{centering}
\epsscale{0.7}
  \plotone{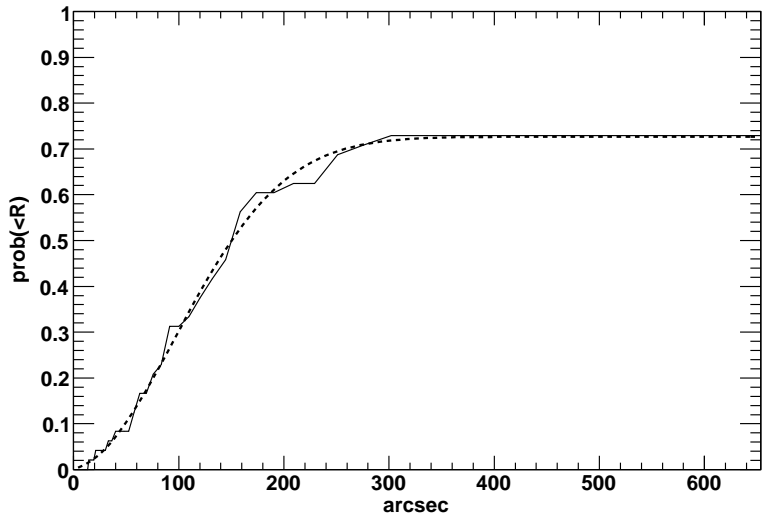}
\caption{Gaussian fit to the curve of Fig. \ref{fig:ros_corr}. 
The PSLA at 95\% and
99\% confidence is respectively 3\hbox{$.\!\!^{\prime}$3} and 5\arcmin.}
	\label{fig:fit_ros}
\end{centering}
\end{figure}

We have detected 49 hard X-ray sources in our survey. Four of these sources are
residuals caused by imperfect modeling (and inclusion in the background model)
of bright sources which are detected in individual DPHs. 
These 4 sources (LMC~X--4, EXO~0748$-$676, Vel~X$-$1 and V$^*$~V1055~Ori)
are still detected in the reconstructed image with a S/N of 20--40.
\\\\
In table~\ref{tab:src}, we report the coordinates and fluxes of all
45  serendipitous objects detected above the 4.5\,$\sigma$  
detection threshold.
\\\\
We have correlated our sources with the ROSAT All-Sky Survey Bright Source Catalogue
\citep{voges99} in the same way as in \cite{stephen06}. 
In Fig.~\ref{fig:ros_corr}, we report  the number of BAT sources which have at least
one ROSAT source within a given radius. Also, to understand the contribution 
of chance coincidences to these associations, we  
performed a Monte Carlo simulation using 5$\times 10^5$ 
positions randomly distributed in our
field. Due to non-uniformity  in the distribution of ROSAT sources, the probability
of a chance association increases  slightly towards negative 
Galactic latitudes. Taking into account the highest density of ROSAT sources
(for $-40^{\circ}<$b$<-20^{\circ}$), we get from Fig.~\ref{fig:ros_corr}
that using a radius of 300\arcsec\ for the identification
of our sources will yield a probability of chance coincidence of 0.015 
(1 wrong identification overall).
The same figure yields
also information about the BAT point spread function location accuracy (PSLA),
as the BAT uncertainty in the position dominates the ROSAT error. Thus, 
assuming that the ROSAT position is the ``true'' source position and
considering only the ROSAT associations, we fitted an inverted Gaussian to the curve
of Fig.~\ref{fig:ros_corr} (see Fig.~\ref{fig:fit_ros}); we
derived that 95\% and 99\% of all spatial coincidences
are within 3\hbox{$.\!\!^{\prime}$3}  and 5\arcmin\ respectively. 
Thus, using a 5\arcmin\ radius for source identification yields the 
best compromise between probability of finding the BAT counterpart and
chance coincidence.

\begin{figure}[h!]
\begin{centering}
\epsscale{0.7}
\plotone{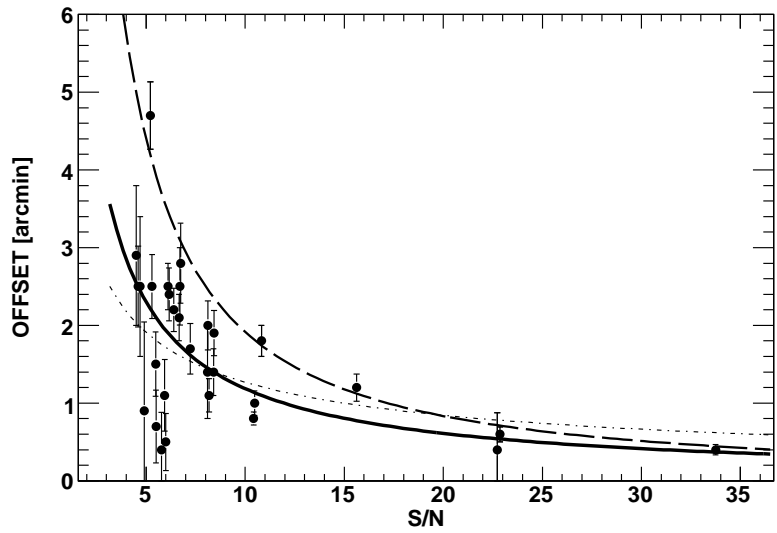}
\caption{Offset of the  sources detected in the final survey image
as a function of their significance.
The solid line is the fit to the data with the function 
OFFSET = 10\hbox{$.\!\!^{\prime}$7}$\times \textup{S/N}^{-0.95}$.
The long dashed line is the fit
to $>3$\,$\sigma$ deviations from the previous fit and gives the maximum expected
offset for a given significance. For comparison, the short dashed line shows
the best fit (Eq. \ref{eq:off}) to the offset-significance relation for
 sources detected in individual pointings.}
	\label{fig:snr_off}
\end{centering}
\end{figure}

It is not surprising that $<$70\% of our
sample is correlated with the ROSAT catalog since photoelectric absorption might
play an important role.
Using the ROSAT catalog we achieved to identify 30 of our sources.
These sources are generally the brightest of our sample and they
were already detected by previous observatories \citep{macomb99}.
\\\\
Using the same 300\arcsec\ error radius, we searched for spatial coincidences
between our sources and both 
the HEAO-1 catalog of high energy sources \citep{levine84}
and  the 2nd INTEGRAL-IBIS catalog \citep{bird06}.
We found that 2 sources were already detected in hard X-rays by HEAO-1
and 7 objects, including also the previous 2, by INTEGRAL. 
All these 7 sources  were already detected at low energy by ROSAT. 
Two additional sources, 3C 227 and V* BG CMi have an Einstein IPC counterpart 
\citep{harris90}. 3C 227 was also detected during a  long (11\,ks) ROSAT-PSPC
observations \citep{crawford95}. 
\\\\
Some of the new sources can be identified using the narrow field X-ray telescope (XRT)
on board Swift. With its 5\arcsec\ position 
accuracy XRT is able to pinpoint the source counterpart in less than 2 ks.
We requested and obtained 3 followup observations of our targets 
(J0732.5-1330, J0823.3-0456 and J0918.6+1617) and this
allowed us to firmly identify the counterpart of those sources \citep{ajello06a}.
Other sources (e.g. J0916.4-6221, J0519.5-3240, J0505.8-2351 and J0920.8-0805) 
were observed
by XRT as part of the ongoing effort of the BAT all-sky survey 
\citep{tueller05a,tueller05b,kennea05atel3}.
We also searched the Swift archive for XRT observations covering
the fields of our sources. 
\\
A total of 17 sources can be firmly identified thanks to XRT.
The results of all the identification efforts using X-ray catalogs and XRT, are
reported in table \ref{tab:src}.
Details of all sources identified using XRT are  
given case-by-case in the next section.
\\
Using the sources with a known X-ray counterpart, 
we report, in Fig. \ref{fig:snr_off}, the sources' offsets
from their catalog position as a function of significance. A fit to the 
data shows that the offset varies with S/N as:
\begin{equation}
\textup{ OFFSET} = 10.7(\pm 1.9)\times \textup{S/N}^{-0.95 (\pm0.08)}\ \ \ \ [\textup{arcmin}] 
\label{eq:off_survey}
\end{equation}
Moreover, from the same plot we expect that for a 4.5\,$\sigma$
detection the maximum  offset be 5\arcmin; this is in perfect agreement
with what is shown in Fig. \ref{fig:fit_ros}.\\
The offset derived for a 10\,$\sigma$ source from the previous relation
and from equation \ref{eq:off} (i.e. the same 10\,$\sigma$ source is detected
in the single 300\,s sky image) is 
1\hbox{$.\!\!^{\prime}$18} and  1\hbox{$.\!\!^{\prime}$26} respectively.
The small difference between the per-pointing location accuracy
and the accuracy in the summed image is due to the fact that Equations
\ref{eq:off} and \ref{eq:off_survey} are computed for different ranges
of S/N. Indeed, sources detected in the survey image (sum of 2671 shorter
observations) span the 4.5--10 range of significance while most of the sources
detected in single pointings have S/N greater than 9\,$\sigma$ 
(see Fig.~\ref{fig:off5min_sn}).
Thus, we can affirm that our survey analysis preserves 
the good location accuracy of BAT.
%%%%%%
%%%%%%

%%%%%%%%%%%%%%%%%%%%%%%%%%%%%%%%%%%%%%%%%%%%%%%%%%%%%%%%%%%%%%%%%%%%%%%%%%%%%%
%---------XRT observations
%
\subsection{XRT observations}
{\bf SWIFT J0407.6+0336.} XRT observed this source field for 7\,ks on Jul 11, 2006.
The only detected object, RA(2000)= 04 07 16.2 Dec(2000)=+03 42 24.3,
is coincident with the Sy2 galaxy 3C 105.0 and distant 
4\hbox{$.\!\!^{\prime}$7} from
the BAT position. It is the first time that 3C 105.0 is detected in X-rays.
\\\\
{\bf SWIFT J0505.8-2351.} XRT observed this source field for 3.2\,ks on Aug. 20, 2005.
Only one source is detected within the BAT error box 
at  RA(2000)=05 05 45.4 Dec(2000)=-23 51 16.8 coincident with the Sy2 galaxy
2MASX J05054575-2351139. This object was already identified as the BAT counterpart
in \cite{tueller05b}.
\\\\
{\bf SWIFT J0519.5-3240.} A 7\,ks XRT observation was performed on Nov. 26, 2005. In the XRT
field only two objects are detected. The brighter one at  
RA(2000)=05 19 35.5 Dec(2000)=-32 39 22.4 is only 
1\hbox{$.\!\!^{\prime}$1} from the BAT position.
The fainter one is detected at RA(2000)=05 19 25.8 Dec(2000)=-32 42 32.3
and it is only a 2$\sigma$ detection. The bright source is associated with 
the nearby Sy1.5 galaxy ESO 362- G 018 (also detected by ROSAT
as 1RXS J051936.1-323910). 
ESO 362- G 018 was already identified as the BAT
counterpart by \cite{tueller05b}.
\\\\
{\bf SWIFT J0522.6-3625.} XRT observed this field for 899\,s in May 26 2005. Only one
source is detected at RA(2000)=05 22 57.8 Dec(2000)=-36 27 29.7 at 
4\hbox{$.\!\!^{\prime}$6}
from the BAT position. The XRT source is coincident with ESO 362-G021 a BL Lac
object already detected by ROSAT and XMM at lower energies and by BeppoSAX
in hard X-rays \citep{donato05}.
\\\\
{\bf SWIFT J0539.9-2842.} An XRT observation of 14\,ks 
took place on Dec. 8, 2005.
A faint source is detected at RA(2000)=05 39 09.3 Dec(2000)=-28 41 01.5
coincident with the z=3.1 QSO PKS 0537-2843 and distant 
2\hbox{$.\!\!^{\prime}$5} from the BAT
position. The QSO  was discovered in X-rays by the Einstein observatory 
 \citep{zamorani81}
and then studied by ROSAT, ASCA and lately by XMM. The BAT detection in hard 
X-rays is the first to date.
\\\\
{\bf SWIFT J0550.8-3215.} A 9\,ks XRT observation took place on May 21, 2005. 
A very bright source is detected at RA(2000)=05 50 40.4 Dec(2000)=-32 16 15.5 
distant  2\hbox{$.\!\!^{\prime}$4} from the BAT position. 
The XRT source is associated
with a well known blazar PKS 0548-322 already detected in hard X-ray
\citep[see][]{donato05}. The blazar is then the BAT counterpart.
\\\\
{\bf  SWIFT J0552.1-0727.} During 9\,ks of exposure on Apr. 8, 2006, XRT detects
a bright source located at RA(2000)=05 52 11.5 Dec(2000)=-07 27 24.2.
The object is coincident with the well known Sy2 galaxy NGC 2110 and 
its position is only 0\hbox{$.\!\!^{\prime}$4} away from the BAT detection.
The detection of NGC 2110 in the 3-20\,keV band by  RXTE  \citep{revnivtsev04}
and the presence of no other source in the XRT field secure the identification
of NGC 2110 as the BAT counterpart.
\\\\
{\bf SWIFT J0640.0-2553.} During 2.8\,ks of observation on Mar. 23, 2006,
XRT detects only one bright source at RA(2000)=06 40 11.8 Dec(2000)=-25 53 41.5
coincident with the Sy1 galaxy ESO 490- G26 (already detected in soft X-rays by 
ROSAT as RX J064011-25536). 
The source position is 2\hbox{$.\!\!^{\prime}$5} distant from the BAT detection.
\\\\
{\bf SWIFT J0732.5-1330, aka SWIFT J0732.5-1331.} 
XRT observed this field for 4\,ks on Apr. 28, 2006. Only
one source is detected in the XRT field at  RA(2000)=07 32 37.7 
Dec(2000)=-13 31 08.6. This source is coincident with an 
USNO B1 star J073237.64-133109.0. The source was already identified as
the BAT counterpart by \cite{ajello06a}. Follow-up measurements 
in the optical determined that this source is a new intermediate polar 
\citep[][and references therein]{weathley06}.
\\\\
{\bf SWIFT J0759.9-3844.} XRT observed the  field of this source for 7\,ks. 
Three sources
are clearly detected. The brightest of them is located at RA(2000)=07 59 41.2 
Dec(2000)=-38 43 57.9 being only  
0\hbox{$.\!\!^{\prime}$5}
away from the BAT position while
the remaining two are distant more than 10\arcmin.
The brightest object is coincident with the INTEGRAL source IGR J07597-3842 
and with the ROSAT source RX J075942.0 -384359.
The fact that the only source within 4\arcmin\ from the BAT position is also
detected by INTEGRAL in hard X-rays \citep{denhartog04} 
makes  this source the BAT counterpart.
The probable association with  an IR point source 
IRAS 07579-3835, and a 1.4-GHz radio counterpart (NVSS archive) makes the case
for a AGN nature of the object. This source was  identified as
being a Sy1.2 galaxy during a recent optical spectroscopy followup \citep{masetti06b}.
\\\\
{\bf SWIFT J0823.3-0456.} Only a single faint source is detected by XRT 
during 1.2\,ks of
exposure on Jan. 6, 2006. The source is located at 
RA(2000)=08 23 01   Dec(2000)=-04 56 02.5 and 2\arcmin\ away 
from the BAT detection. 
The object is coincident with the galaxy FAIRALL 0272 and 
was already identified as the BAT counterpart by \cite{ajello06a}.
An optical follow-up showed that the source is a  Sy2 galaxy \citep{masetti06}.
\\\\
{\bf SWIFT J0918.6+1617.} SWIFT J0918.5+1617, aka SWIFT J0918.5+1618,
 is another source found thanks to our algorithm \citep{ajello06a}.
During an XRT followup of 0.6\,ks, the only detected source
is located at RA(2000)=09 18 25.8 Dec(2000)=+16 18 20.8  
(2\hbox{$.\!\!^{\prime}$5} away
from the BAT position) and coincident with the galaxy Mrk 704.
Mrk 704 was previously detected in soft X-rays by ROSAT \citep{schwope00}.
In a recent optical followup, the galaxy was found to be a Sy1 \citep{masetti06}.
\\\\
{\bf SWIFT J0920.8-0805.} An XRT observation of 8.5\,ks took place on Dec. 10, 2005. 
Only one source is detected in the entire field. Its position, 
RA(2000)=09 20 46.0 Dec(2000)=-08 03 21.8, is coincident with
the Sy2 galaxy MCG-01-24-012 and distant 
2\hbox{$.\!\!^{\prime}$2}
from the BAT position.
This object was already identified as the BAT counterpart by \cite{kennea05atel3}.
\\\\
{\bf  SWIFT J0945.9-1421.} An XRT observation of 11\,ks took place on Jul. 8, 2006.
The only  source detected inside the BAT error box
is located at RA(2000)=09 45 42.0 Dec(2000)=-14 19 33.7. The source
is coincident with the Sy1.9 galaxy NGC 2992 and is distant
1\hbox{$.\!\!^{\prime}$5}
from the BAT detection. The source was already detected in soft X-rays
by ROSAT as 1RXS J094541.9-141927.
\\\\
{\bf SWIFT J0947.6-3056.} XRT observed the source field for 10\,ks on Dec. 9, 2005.
Only one bright source is detected at RA(2000)=09 47 39.8 Dec(2000)=-30 56 55.4
coincident with the Sy2 galaxy ESO 434- G 040 and distant only 
0\hbox{$.\!\!^{\prime}$4}
from the BAT position. The galaxy was also detected in hard X-ray by INTEGRAL
\citep{bird06}.

%%%%%%%%%%%%%%%%%%%%%%%%%%%%%%%%%%%%%%%%%%%%%%%%%%%%%%%%%%%%%%%%%%%%%%%%%%%%
%
%         CONCLUSION
%
\section{Conclusions}

\begin{figure*}[ht!]
  \begin{center}
  \begin{tabular}{cc}
  \includegraphics[scale=0.41]{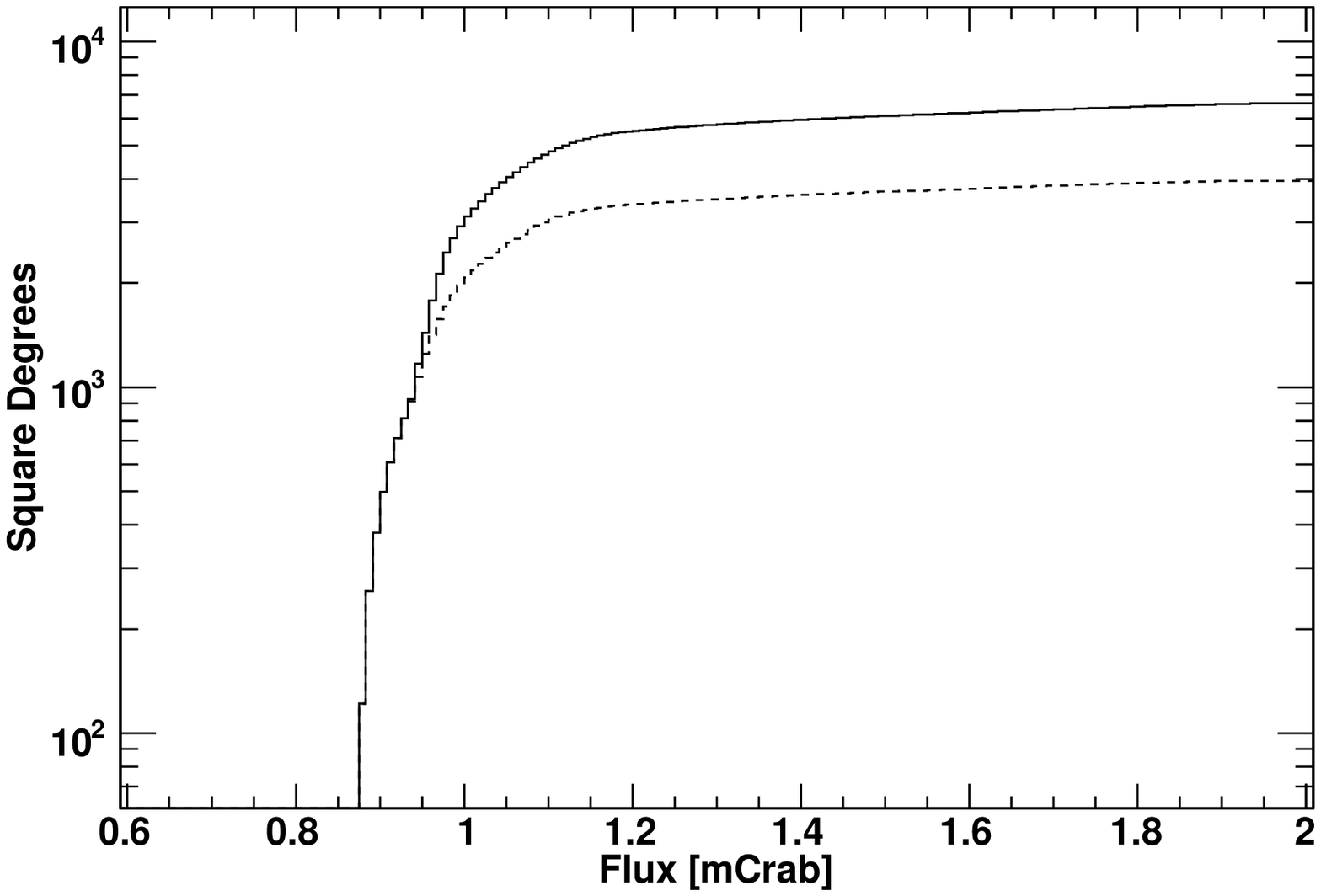} 
	 \includegraphics[scale=0.41]{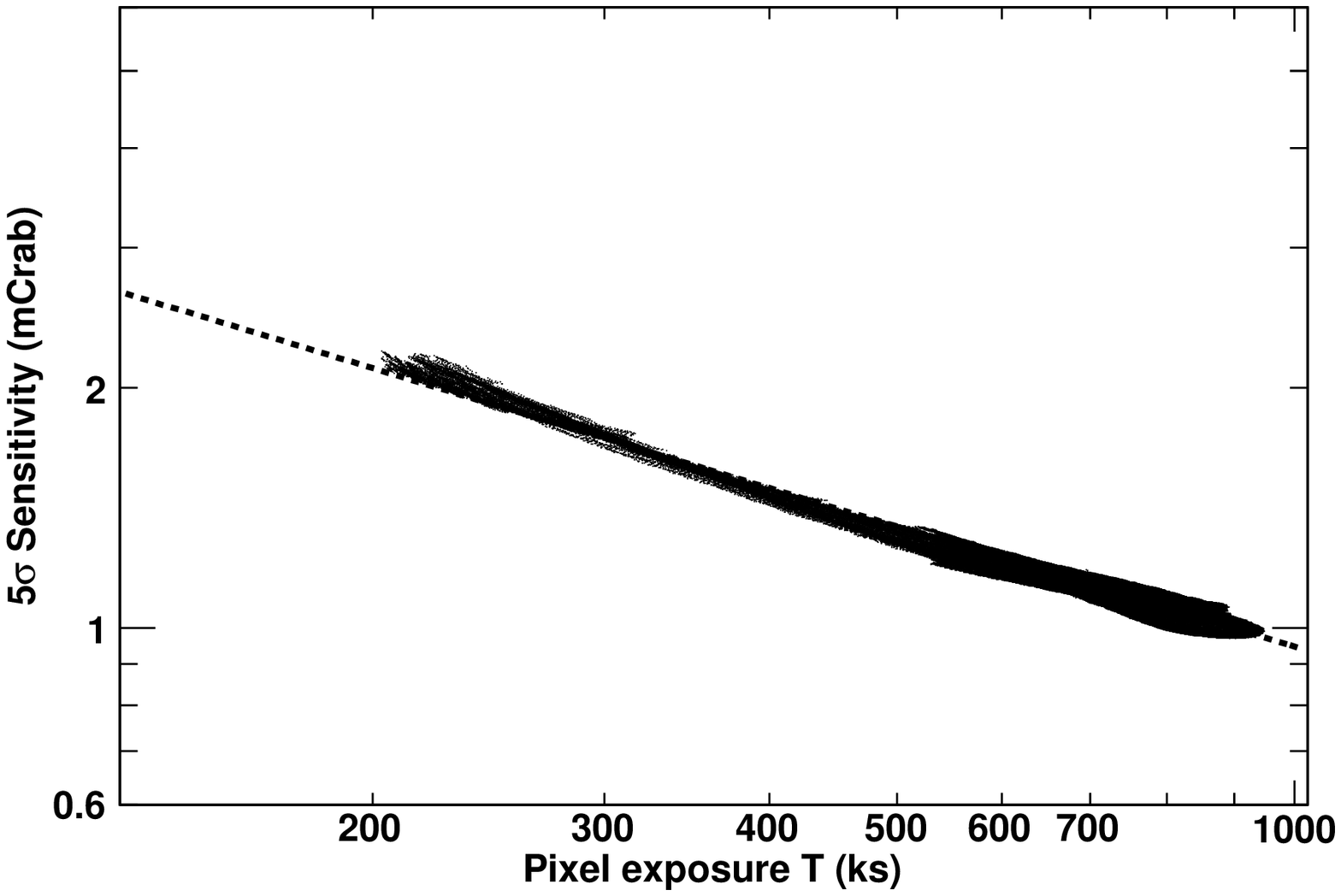}\\
\end{tabular}
  \end{center}
  \caption{
{\bf Left Panel:}
Sky coverage as a function of minimum detectable flux for S/N=4.5
in the 14-170\,keV band. As it can be seen 75\% of the surveyed area is covered
to 1\,mCrab. The dashed  line is the sky coverage for $|b|>15^{\circ}$. 
{\bf Right Panel:}
Pixel 5\,$\sigma$ sensitivity threshold as a function of pixel exposure time. 
The dashed line is the fit to data points  corresponding to the 
function 3.0mCrab$(T/100\textup{ks})^{-0.5}$.
}
  \label{fig:noise}
\end{figure*}

We have presented an application of the Maximum Likelihood method as a deconvolution
technique used to reconstruct the sky image when dealing with a coded-mask instrument
like BAT. The main difference with other image reconstruction algorithms, such as the 
standard cross-correlation technique, is that a sky distribution model
is forward-folded through the full instrumental response and fit to
the detector plane counts in order to derive the most probable sky image. 
This is realized in a single step 
including data from many pointings and thus no image mosaicking is  required.
This study was motivated principally by the capabilities of ML to: 1) 
preserve the full statistical information in the data and 2) to reduce the systematic
errors connected to mosaicking techniques which other methods cannot avoid. 
This leads to an improvement in sensitivity over other methods.
\\\\
Moreover, this study is motivated by the need to use sensitive
imaging techniques for the study of the hard X-ray sky. Although deep
soft  X-ray surveys (0.5-2.0\,keV) were able to resolve the majority
of the CXB emission into discrete sources \citep{hasinger98}, only a minor 
fraction of the CXB above 8\,keV  is resolved \citep{worsley05}. 
Furthermore, the bulk of the CXB radiation is emitted around
30\,keV \citep{marshall80} and the exact nature of the source population
responsible for the background at these energies is unknown because of the 
low sensitivity of previous hard X-ray telescopes. 
The BAT
coded mask detector, on board the Swift mission, represents a major 
improvement in sensitivity for imaging of the hard X-ray sky; thus,
we tested  our ML imaging algorithm on BAT survey data.
This study was also complemented by an optical spectroscopy
campaign aimed at identifying BAT-discovered extragalactic hard X-ray objects
\citep{rau07}.
\\\\
The results presented in the previous sections can be summarized as follows:
after screening our dataset for bad data as discussed in
Section \ref{sec:preproc}, the final survey image 
obtained using the ML method presents a perfect Gaussian normal
noise.  We detected 49 hard X-ray sources  
above the  4.5\,$\sigma$ detection threshold. 
Only 12 were previously
known as hard X-ray emitters (previously detected by INTEGRAL or HEAO-1).
37 are new sources detected by BAT due to our  image reconstruction method.\\
The correlation of BAT sources with the ROSAT catalog shows the extremely good
location accuracy of the BAT instrument which is also preserved by our algorithm.
Also it is worth noticing that $\sim$30\%
of our sources  are not correlated with
the ROSAT objects; this is most probably due to the presence of 
photoelectric absorption in some of the new BAT  sources.
The analysis of the limiting flux as a function of pixel exposure 
(see Fig.~\ref{fig:noise})
for the reconstructed image sum of all observations,
shows that systematic errors do not dominate over statistical ones and
that BAT should be able to achieve, in the future, 
a sensitivity of 0.5\,mCrab with 3\,Ms 
of exposure (if systematics remain at this level).
The sky coverage 
shows that 75\% of the survey
is covered to flux $\sim$1\,mCrab and all of it to fluxes $>2.0$\,mCrab. 
All of this makes this analysis one of the most sensitive surveys ever performed
in the hard X-ray domain.
\\\\
The optical spectroscopy identification of the new sources and 
a discussion about the optical properties are left to a separate paper
\citep{rau07} while the statistical and spectral X-ray properties
will be discussed in \cite{ajello07b}.

\acknowledgments
MA acknowledges N. Gehrels and the BAT team for the warm
hospitality and enlightening discussions,
A. Yoldas for all his tips\&tricks about parallel programming and  python.
The anonymous referee is aknowledged for his helpful comments which
improved the manuscript.
This research has made use of the NASA/IPAC Extragalactic Database (NED) which
is operated by the Jet Propulsion Laboratory, of data obtained from the 
High Energy Astrophysics Science Archive Research Center (HEASARC) provided 
by NASA's Goddard Space Flight Center, of the SIMBAD Astronomical Database
which is operated by the Centre de Donn\'ees astronomiques de Strasbourg, of
the Sloan Digital Sky Survey (SDSS) managed by the Astrophysical Research 
Consortium (ARC) for the Participating Institutions and of the ROSAT All Sky
Survey mantained by the Max Planck Institut f\"ur extraterrestrische Physik.

\bibliographystyle{apj}
\bibliography{/Users/marcoajello/Work/Papers/BiblioLib/biblio}

\clearpage

%%%%%%%%%%%%%%%%%%%%%%%%%% Tables

\begin{deluxetable}{lr}
\tablewidth{0pt}
\tablecaption{Fraction  of rejected data due to any single criterion
\label{tab:data_frac}}
\tablehead{
\colhead{TYPE}    & \colhead{FRACTION}    
}
\startdata
Star tracker lock   & 9.7\% \\
Pointing stability  & 17.4\% \\
Outside SAA         & 10\%   \\
Exposure $>$ 200 s  & 7\%    \\
Others              & $<$1 \% \\
All conditions      & 34   \% \\

\enddata
%\tablenotetext{a}{}
\end{deluxetable}

%\clearpage

\begin{deluxetable}{rrccccccc}
%\tabletypesize{\scriptsize}
\tablewidth{0pt}
\tablecaption{Sources detected during data screening\label{tab:prec_src}}
\tablehead{
\colhead{RA}             & \colhead{DEC}     & \colhead{S/N(max)}    &
\colhead{TOTAL S/N}      & \colhead{\# detections}  & \colhead{Type}        &
\colhead{ID}             \\
%%%%%%%%%  units
\colhead{\scriptsize (J2000)}  & \colhead{\scriptsize (J2000)}    

}
\startdata
%RA & DEC & SNR(max) &  TOTAL SNR & NofDet &  ID  \\
19.1746  & -73.4559 & 16.9  & 47.7  & 22   & HXB  & SMC X-1        \\
58.8665  & +31.0270 & 8.3   & 8.3   & 1     & HXB  & V* X Per     \\
83.2654  & -66.3567 & 11.5  & 35.0  & 17   & HXB  & LMC X-4        \\
83.6265  & +22.0079 & 126.9 & 1505  & 581  & PSR  & Crab           \\
84.7121  & +26.2949 & 90.5  & 195   & 61  & HXB  & 1A 0535+26     \\
94.2964  & +9.1199  & 8.7   & 12.6  & 3    & LXB  & V* V1055 Ori   \\
117.2030 & -67.7483 & 8.0   & 30.0  & 15   & LXB  &  EXO 0748-676  \\
135.5234 & -40.5565 & 139.3 & 917   & 888  & HXB  & Vel X-1        \\
152.4996 & -58.2222 & 13.7  & 38.8  & 18   & HXB  & GRO J1008-57   \\
170.2396 & -60.5556 & 50.2  & 161.7 & 79   & HXB  & Cen X-3        \\
176.9183 & -61.9794 & 8.9   & 8.9   & 1     & HXB  & V* V830 Cen   \\
186.6980 & -62.7649 & 98.0  & 401   & 284  & HXB  & GX 301-2       \\
201.4286 & -43.0056 & 8.8   & 12.6  & 2    & Sy2  & Cen A          \\
235.5705 & -52.3704 & 12.2  & 14.4  & 2    & HXB  & V* QV Nor      \\
243.0916 & -52.4028 & 16.1  & 40.7  & 19   & LXB  & H 1608-522     \\
244.9860 & -15.6526 & 25.5  & 76.0  & 24   & LXB  & Sco X-1        \\
247.9922 & -48.8173 &  8.8  & 11.5  & 2    & HXB  & IGR J16318-4848\\
250.2720 & -53.7497 & 8.9   & 19.6  & 6    & LXB  & H 1636-536     \\
251.4481 & -45.6088 & 13.1  & 21.8  & 6    & LXB  & GX 340+0       \\
255.1860 & -41.6435 & 27.0  & 38.9  & 8    & HXB  & EXO 1657-419   \\
255.9813 & -37.8334 & 50.2  & 98.3  & 22   & HXB  & V* V884 Sco    \\
256.4394 & -36.4345 & 12.1  & 13.2  & 2    & LXB  & Sco X-2        \\

%186.5026 & -62.7640 & 11.1  & & & HXB & V* BP Cru \\ maybe GX 301-2@ wrong position

\enddata
\tablenotetext{a}{Note that the significance of the detection depends on the
source intensity, exposure and  on the position in the FOV. The ``TOTAL S/N''
was computed as the sum of the squared significances in each observation.
%The ``flag'' column shows which source is also detected at some residual 
%significance ``S/N$_{resid}$'' in the co-added image.
}
%% You can append references to a table using the \tablerefs command.
%\tablerefs{If we want ot put some ref here.}
\end{deluxetable}

\begin{deluxetable}{llcccc}
\tablewidth{0pt}
\tablecaption{Sensitivities of different hard X-ray surveys.\label{tab:sens}}
\tablehead{
\colhead{Instrument}    & \colhead{Ref.} & \colhead{Energy}&
\colhead{S/N$_{TH}$} & 
\colhead{Sensitivity\tablenotemark{a}}   & \colhead{Equivalent 5\,$\sigma$ Sens\tablenotemark{b}} \\  
\colhead{}& \colhead{}&\colhead{\scriptsize{[keV]}} & \colhead{} & \colhead{\scriptsize(@1Ms [mCrab])}& 
\colhead{\scriptsize(@1Ms [mCrab])}
}
\startdata
INTEGRAL  & Bassani  et al., 2006  &20--100  & 5.0   & 0.8   & 0.8 \\ 
BAT       & Markwardt et al., 2005 &15--195  & 5.5   & 1.3   & 1.2\tablenotemark{c}  \\
BAT       & this work              &14--170  & 4.5   & 0.86  & 0.95 \\
\enddata
\tablenotetext{a}{The sensitivity is computed by considering the noise-exposure
relation provided by the authors (e.g. right panel of  Fig.
\ref{fig:noise}) evaluated at  the threshold S/N they used to detect 
sources.}
\tablenotetext{b}{Sensitivities are computed assuming a threshold S/N of 
5\,$\sigma$ for all  instruments/surveys.}
\tablenotetext{c}{The sensitivity reported by \cite{markwardt05} is referred
to the all-sky analysis. Their best 5\,$\sigma$ 
sensitivity in high-latitude fields is $\sim$1.0\,mCrab for 1\,Ms of exposure.}
\end{deluxetable}

%%%%%%%%%%%%%%%%%%%%%%% this is the main table
\clearpage
\begin{deluxetable}{lccccclccll}
\tablewidth{0pt}
\tabletypesize{\scriptsize}
\rotate
\tablecaption{Detected hard X-ray sources \label{tab:src}}
\tablehead{
%%%%%%%% column names
\colhead{SWIFT NAME}               & \colhead{R.A.}       &
\colhead{DEC}            & \colhead{Flux}     &
\colhead{S/N}            & \colhead{Exposure}    & \colhead{ID} & 
\colhead{Offset}         & \colhead{Type}     & 
\colhead{Catalogs\tablenotemark{a}  }     & \colhead{XRT position (ref.)} \\
%%%%%%%%%  units
\colhead{}               & \colhead{\scriptsize (J2000)}  &
\colhead{\scriptsize (J2000)}       & \colhead{\scriptsize(10$^{-11}$ cgs)} &
\colhead{}               & \colhead{\scriptsize{(100 ks) }}         & \colhead{} &
\colhead{\scriptsize(arcmin)}
}
\startdata

J0407.6+0336& 61.9178 & 3.6517 & 4.21  & 5.2  & 330 & 3C 105.0   & 4.7 
  & Sy2 & X  & 04 07 16.2, +03 42 23 \\

J0426.4-5710& 66.6021 & -57.1775 & 2.02  & 4.5 & 803 & 1AXG J042556-5711\tablenotemark{b} & 3.5
 & Sy1 & R  & \nodata \\

J0433.1+0520& 68.2982 & 5.3374 & 7.0  & 10.4  & 316 &  3C 120\tablenotemark{b}  & 1
  & Sy1 & R  & \nodata \\

J0451.6-0349& 72.9205 & -3.8240 & 3.73  & 4.9 & 416 & MCG -01-13-025\tablenotemark{b} & 0.9
  & Sy1.2 & R  & \nodata \\

J0505.8-2351 & 76.4674 & -23.8666 &  2.62  & 7.2 & 816 &   SWIFT J0505.7-2348 & 1.7
  & Sy2 & X  & 05 05 45.4 -23 51 17.0 (1) \\

J0510.8+1631& 77.7224 & 16.5265 & 7.37  & 4.7 & 231  &  CSV 6150\tablenotemark{b}   & 2.5
 & Sy1.5 & R  & \nodata \\

J0514.0-4003 & 78.5146 & -40.0558 & 5.02  & 10.4  &934 & 4U 0513-40 \tablenotemark{b}   & 0.8 
  & LXB & R  & \nodata \\

J0516.0-0007& 79.0096 & -0.1332 & 7.6  & 6.1  & 373 &  QSO B0513-002\tablenotemark{b}  & 2.5
  & QSO-Sy1 & R  & \nodata \\

J0517.1+1633& 79.2839 & 16.5605 & 3.80  & 4.5  &  233  & \nodata &\nodata
 & \nodata & \nodata  & \nodata \\

J0519.5-3240 & 79.8844 & -32.6720 & 4.78  & 8.2  & 909 &  ESO 362- G 018\tablenotemark{b}  & 1.1
  & Sy1.5 & R, X  &  05 19 35.5, -32 39 22 (1)\\

J0519.7-4545 & 79.9460 & -45.7557 & 4.48  & 5.3  & 904 & Pictor A\tablenotemark{b} & 1.5 
  & Sy1 & R  & \nodata \\

J0522.6-3625 & 80.6581  &  -36.4233    & 3.65 & 4.7 &927  &  ESO 362-G021\tablenotemark{b} & 4.6
  & BLLAC &   R, X  & 05 22 57.9, -36 27 29\\

J0529.4-3247 & 82.3541 & -32.7965 & 5.15  & 10.9 & 904  &  V* TV Col\tablenotemark{b} & 1.3
  & CV-DQ* & R  & \nodata \\

J0534.5-5801& 83.6470 & -58.0200 & 1.8 & 5.6 & 800 & V* TW Pic\tablenotemark{b}   & 2.1
  & CV & R, I & \nodata \\

J0539.0-6406 & 84.7717 & -64.1148 & 2.10 & 6.6 &  726  & LMC X-3\tablenotemark{b} & 2.1 
  & HXB & R  & \nodata \\

J0539.5-6943 & 84.8917 & -69.7210 & 4.48 & 8.4 &  681  &LMC X-1\tablenotemark{b} & 1.4 
  & HXB & R, I  & \nodata \\

J0539.9-6919 & 84.9878 & -69.3230 & 6.30 & 8.1  & 681 & PSR B0540-69.3\tablenotemark{b} & 1.4 
  & Pulsar & R, I  & \nodata \\

J0539.9-2842 & 84.9953 & -28.7029 & 5.68  & 4.6  &858  &  PKS 0537-286\tablenotemark{b} &2.5
  & Blazar & R, X  & 05 39 54.1, -28 39 54 \\

J0550.8-3215 & 87.7165 & -32.2610 & 3.40  & 6.1 &   883 & PKS 0548-322\tablenotemark{b} & 2.4
  & BL Lac & R, X  & 05 50 40.4, -32 16 15 \\

J0552.1-0727& 88.0411 & -7.4554 & 32.04  &  33.7 & 477 & NGC 2110   & 0.4
  & Sy2 & X  & 05 52 11.5  -07 27 24  \\

J0558.0-3822 & 89.5237 & -38.3799 & 3.62  & 6.7 & 913 &  LEDA 75476\tablenotemark{b}  & 2.8
  & Sy1 & R  & \nodata \\

J0640.0-2553 & 100.0031 & -25.8931 & 2.54  & 5.3 & 793 &  ESO 490- G 26\tablenotemark{b}  & 2.5
  & Sy1.2 & R, X & 06 40 11.8, 25 53 41 \\

J0727.5-2406 & 111.8951 & -24.1039 & 1.9  & 5.3  & 850 &  \nodata  &\nodata
  & \nodata & R\tablenotemark{e}  & \nodata \\

J0728.6-2604& 112.1626 & -26.0696 &1.87 & 4.7 & 851  & V* V441 Pup\tablenotemark{b} & 3.5
 & HXB & R & \nodata \\

J0731.5+0955& 112.8752 & 9.9214 & 3.30  & 5.9  & 578   &  V* BG CMi  & 1.3
  & CV & E  & \nodata \\

J0732.5-1330& 113.1328 & -13.5037 & 1.85  & 5.9 & 820 & SWIFT J0732.5-1331  & 1.1
  & CV & X & 07 32 37.8, -13 31 07 (3) \\

J0739.6-3144 & 114.9127 & -31.7496 & 2.77  & 6.2  & 846 &  \nodata &\nodata
  & \nodata & \nodata  &  \nodata \\

J0743.0-2543 & 115.7501   & -25.7314 &  1.86 & 4.95 & 869  &  \nodata &\nodata
  & \nodata & R\tablenotemark{e}  &  \nodata \\

J0759.9-3844 & 119.9822 & -38.7422 & 4.90  & 6.0 & 804  &  IGR J07597-3842  &0.5
  & Sy1.2 & R, I, X  & 07 59 41.2, -38 43 57 \\

J0804.2+0507& 121.0552 & 5.1203 & 4.1 & 8.4 & 736  &  UGC 4203\tablenotemark{b}  
& 1.9   & Sy2 & R, X & 08 04 05.4 +05 06 49\\

J0811.5+0937& 122.8750 & 9.6214 & 2.26  & 4.6  & 700 &  \nodata  & \nodata
  & \nodata & R\tablenotemark{d}  & \nodata \\

J0823.3-0456 & 125.8271 & -4.9401 & 3.23  & 8.1 & 846 & SWIFT J0823.4-0457  & 2.0
  & Sy2 & X  & 08 23 01.0, -04 56 02 (3) \\

J0835.3-4510 & 128.8308 & -45.1771 & 17.1  & 22.7 &705 & Vela PSR\tablenotemark{b} & 0.4 
  & Pulsar & R, I, H  & \nodata \\

J0838.4-3559 & 129.6151 & -35.9976 & 4.01  & 5.5 & 791 &  FRL 1146  &0.7
  & Sy1 & R, I  & \nodata \\

J0839.8-1214& 129.9556 & -12.2467 & 2.43  & 5.8 & 866 & 3C 206\tablenotemark{b}  & 0.4
  & QSO & R  & \nodata \\

 J0844.9-3531 & 131.2411 & -35.5313 & 2.97  & 5.0 & 780  &  \nodata &\nodata
  & \nodata & \nodata  & \nodata \\

J0854.7+1502& 133.6828 & 15.0371 & 5.58  & 5.1 & 646 & \nodata & \nodata
  & \nodata & \nodata  & \nodata \\

J0917.2-6221 &139.112 & -62.359 & 2.07  & 4.5 & 830 & LEDA 90443  & 2.9 
& Sy1 & X & 09 16 08.9, -62 19 29.6 \\

J0918.6+1617& 139.6505 & 16.2987 & 3.3 & 6.7 & 610  &  MRK 0704 & 2.5
  & Sy1.5 & R, X & 09 18 25.9, +16 18 20(3)\\

J0920.3-5512 & 140.0753 & -55.2135 & 9.57  & 15.6  & 612 & 4U 0919-54\tablenotemark{b} & 1.2
  & LXB & R, I, H  & \nodata \\

J0920.8-0805& 140.2134 & -8.0872 & 3.03  & 6.4  &742 & MCG -01-24-012 & 2.2
  & Sy2 & X  & 09 20 46.0, -08 03 21 (2)\\

J0945.9-1421 & 146.4060 & -14.3007 & 5.09  & 5.5 & 621  &  NGC 2992\tablenotemark{b} & 1.5& Sy1.9& R, X  & 09 45 42.1, -14 19 34 \\

J0947.6-3056 & 146.9151 & -30.9388 & 18.01  & 22.8 & 611 &  ESO 434- G 040\tablenotemark{b} &0.6
  & Sy2 & R, I, X  & 09 47 39.8, -30 56 55 \\

J0947.7+0725& 146.9447 & 7.4191 & 5.58  & 4.7   & 581 &  3C 227 & 0.4
  & Sy1 & E,R\tablenotemark{c} & \nodata \\

J0959.5-2251& 149.8805 & -22.8561 & 7.80  & 10.8  & 566 &  NGC 3081\tablenotemark{b}  & 1.8
  & Sy2 & R & \nodata \\

\enddata

\tablenotetext{a}{Catalogs and Instruments used for identification:
R=ROSAT catalog, I= INTEGRAL catalog, H= HEAO-1 catalog, E= Einstein IPC catalog,
X= XRT observations.}

\tablenotetext{b}{Source identified thanks to its proximity to a ROSAT Bright
Source with the method described in the text.}

\tablenotetext{c}{3C 227 was detected in a long (11\,ks) ROSAT-PSPC
observations \citep{crawford95}.}

\tablenotetext{d}{Source J0811.5+0937 has a ROSAT counterpart,
 1WGA J0811.5+0933, reported in the WGA catalog \citep{white94}.
This source coincides with RX J081132.4+093403.}

\tablenotetext{e}{Object has a ROSAT source in the error box, 
however the association has to be proven.}

\tablerefs{ (1) Tueller et al. (2005); (2) Kennea et al. (2005); 
(3) Ajello et al. (2006);
}

\end{deluxetable}

%%%%%%%%%%%%%%%%%%%%%%%%%%%%%%%%%%%%%%%%%%%%%%%%%%%%%%%%%%%%% FIGURES

%%%%%%%%%%%%%%%%%%fig 1
\clearpage

%%%%%%%%%%%%%%%%%%fig 2

%%%%%%%%%%%%%%%%%%fig 3

%%%%%%%%%%%%%%%%%%fig 4

%%%%%%%%%%%%%%%%%%fig 5

%%%%%%%%%%%%%%%%%%fig 6

%%%%%%%%%%%%%%%%%%fig 7
%
%
%
%%%%%%%%%%%%%%%%%%fig 8

%%%%%%%%%%%%%%%%%%fig 9
%
%%%%%%%%%%%%%%%%%%fig 10

\end{document}